**Effects of Irradiation Temperature and Dose Rate on the Mechanical Properties of Self-Ion Implanted Fe and Fe-Cr Alloys**

Christopher D Hardie[a, b], Ceri A Williams[a], Shuo Xu[a] and Steve G Roberts[a]

[a] Department of Materials, University of Oxford, Oxford, OX1 3PH, United Kingdom

[b] EURATOM/CCFE Association, Culham Centre for Fusion Energy (CCFE) Abingdon, Oxfordshire OX14 3DB, UK

| | |
|---|---|
| Name: | Christopher Hardie |
| Address: | Department of Materials, |
| | University of Oxford, |
| | Oxford, |
| | OX1 3PH, |
| | United Kingdom |
| Telephone Number: | +44 (0)1865 273700 |
| Email Address: | christopher.hardie@materials.ox.ac.uk |

# Effects of Irradiation Temperature and Dose Rate on the Mechanical Properties of Self-Ion Implanted Fe and Fe-Cr Alloys

Christopher D Hardie[a, b], Ceri A Williams[a], Shuo Xu[a] and Steve G Roberts[a]

[a] Department of Materials, University of Oxford, Oxford, OX1 3PH, United Kingdom

[b] EURATOM/CCFE Association, Culham Centre for Fusion Energy (CCFE) Abingdon, Oxfordshire OX14 3DB, UK

## ABSTRACT

Pure Fe and model Fe-Cr alloys containing 5, 10 and 14%Cr were irradiated with $Fe^+$ ions at a maximum energy of 2MeV to the same dose of 0.6dpa at temperatures of 300°C, 400°C and 500°C, and at dose rates corresponding to $6 \times 10^{-4}$ dpa/s and $3 \times 10^{-5}$ dpa/s. All materials exhibited an increase in hardness after irradiation at 300°C. After irradiation at 400°C, hardening was observed only in Fe-Cr alloys, and not in the pure Fe. After irradiation at 500°C, no hardening was observed in any of the materials tested. For irradiations at both 300°C and 400°C, greater hardening was found in the Fe-Cr alloys irradiated at the lower dose rate. Transmission electron microscopy and atom probe tomography of Fe 5%Cr identified larger dislocation loop densities and sizes in the alloy irradiated with the high dose rate and Cr precipitation in the alloy irradiated with the low dose rate.

# 1 INTRODUCTION AND BACKGROUND

Ion-implantation enables the production of radiation damage within a matter of hours to doses equivalent to several years in modern nuclear power technologies. This accelerated rate of irradiation is widely considered an advantage for research into the effects of radiation on microstructure and mechanical properties; however, little is known regarding any dose-rate dependence of the characteristics of the radiation damage at a given dose and temperature.

## 1.1 Irradiation Dose Rate

The irradiation dose rate may be expressed in the units of dpa/s; the local density of elementary displacement defects (interstitial – vacancy pairs) per atom (dpa) per unit time. The rate of interaction between these defects is dependent on their mobility and is proportional to the square of their density [1]. The types of interaction and fate of these defects can be classified into three possible reaction paths [2]:

(i) Loss of defects at extended sinks such as dislocations and grain boundaries;

(ii) Growth or shrinkage of defect clusters by the capture of point defects;

(iii) Mutual annihilation by the recombination of a vacancy and interstitial.

At low dose rates and/or high irradiation temperatures, reaction path (i) (sinks) dominates and at a high dose rates and/or low irradiation temperature reaction path (iii) (recombination) dominates [2]. The evolution of radiation damage such as dislocation loops and voids and phenomena such as radiation induced segregation, swelling and creep, depend on the fraction of point defects which migrate to sinks, recombine or cluster within the lattice and will be influenced by the reaction path that dominates the microstructural evolution of the material under irradiation.

The relative proportions of these reaction types are directly dependent on the density and mobility of the defects, and hence dependent on dose rate and temperature. In iron, vacancy type defects are generally found to have significantly higher activation energy for migration compared to interstitial type defects; and defect migration in a material depends strongly on the presence of impurity atoms [3]. In iron the migration energy for a vacancy is 0.67eV and that of an interstitial 0.34eV; carbon forms strongly bound complexes with vacancies and a vacancy-carbon complex has migration energy of 1.08eV [3]. Depending on temperature, this may result in unequal fluxes of mobile interstitials and vacancies, known as a production bias [4], and thus influence the relative fractions of the various reaction paths described above.

Variations in the fraction of reaction paths with dose rate have been inferred from the swelling and creep behaviour of several materials [2, 5, 6]. The majority of research regarding the dose rate dependence of radiation damage was conducted by using fission reactors in the 1980s, which focused on the swelling and creep rates of austenitic stainless steels and their variation with neutron flux. For example, swelling of 316 stainless steel cladding and the creep rate of numerous steels under irradiation has been shown to decrease with increasing dose rate [7, 8]. This decrease in swelling and creep is believed to be due to a reduction in the density of active point defects, resulting from heightened rates of defect clustering or a higher fraction of recombination [6]. Muroga et al. [9] compared the saturated dislocation loop densities in Fe-Cr-Ni austenitic alloys after irradiation with high flux electron, fast neutron and fusion D-T neutron sources, showing a considerable increase in saturated dislocation loop density in the irradiated alloys as the dose rate increased from ~$10^{-9}$ dpa/s (D-T neutron source) to ~$10^{-4}$ dpa/s (electrons). In a subsequent investigation, Fe-15Cr-16Ni irradiated with 4MeV nickel ions at a dose rate of $10^{-4}$ dpa/s exhibited even higher loop densities than those from high energy electron irradiation at a comparable dose rate [10]. This difference may be attributed to heightened rates of defect recombination, resulting from the low PKA energy spectrum characteristic of electron radiation [9] and the corresponding lower rate of production of clusters of defects directly by collision events.

## 1.2  Irradiation Temperature

In addition to the variations found with dose rate, a decrease in dislocation loop density at higher irradiation temperatures can be observed in data from neutron irradiation and from ion implantation [9]; this indicates a higher defect mobility and increased recombination at higher irradiation temperatures. In-situ TEM studies have found similar behaviour with increasing irradiation temperature along with changes in dislocation loop morphology and type [11-13]; however, loop losses to the surface and electron beam-stimulated defect mobility have influenced these results [14]. The increase in irradiation temperature results in higher defect mobility, which causes heightened rates of recombination or annihilation at sinks and accelerated growth/shrinkage of clusters (the three reaction paths discussed in section 1.1).

Rate theory equations have been derived in an attempt to calculate the shift in irradiation temperature required to produce equivalent damage levels for differing dose rates [5, 15]. However, damage evolution in a material depends on multiple reaction processes, each with characteristic activation energies and rates; thus, these equations are usually constricted to a single radiation phenomenon such as swelling.

It has also been observed that the exposure to neutrons during reactor start up and shut down (differing irradiation temperatures), can result in up to 100% difference in final defect structure for a given neutron fluence in some cases [16]. This suggests that microstructural evolution is influenced by the pre-existing radiation damage microstructure, such as the density of stable defects, which effectively act as sinks and may shrink or grow by the absorption of mobile defects. Therefore, in addition to the initial composition and microstructure of a material, it is also important to take the irradiation history into account when predicting the effects of subsequent radiation damage.

This paper reports a study of the irradiation hardening response of Fe and Fe-Cr alloys, subjected to $Fe^+$ irradiation to a dose of 0.6dpa at temperatures of 300°C, 400°C and 500°C and dose rates of $6 \times 10^{-4}$ dpa/s and $3 \times 10^{-5}$ dpa/s. Irradiation hardening of each alloy was characterised by nanoindentation. For material irradiated at 300°C, tests were carried out within single grains to study any influence of crystallographic orientation on the results. Transmission electron microscopy (TEM) and atom probe tomography (APT) of selected Fe 5%Cr samples were conducted to investigate the radiation damage microstructure and Cr distribution after irradiation at the different dose rates.

## 2    METHODS AND MATERIALS

### 2.1    Materials and Preparation

Ultra High Purity polycrystalline Fe and Fe-Cr alloys containing 5, 10 & 14%Cr were produced by EFDA under contract EFDA-06-1901. Table 1 shows their impurity content after hot forming and in the delivered state [17]. All specimens were annealed at temperatures up to 830°C for up to 24 hours in a vacuum $<10^{-4}$ Pa, to produce large grains (~100μm) and remove cold work deformation from the manufacture of the materials. Specimens were lapped (final stage: FEPA P4000 SiC abrasive paper) followed by a chemo-mechanical polish using a suspension of colloidal silica (0.05μm) to remove polishing damage from previous steps.

## 2.2 Ion Implantation

Ion implantation of the materials was performed at the Surrey Ion Beam Centre in the UK. All specimens were mounted flat against a base plate using a custom built sample holder and fixed into position against a heating element. The temperature was controlled and measured by a thermocouple embedded in the sample clamp alongside the specimens. The Monte Carlo code TRIM was used to calculate the damage distribution [18]. These calculations used an average of 1000 ions in pure iron and the recommended displacement energy of 40eV [19]. Chromium has the same displacement energy as iron and has minimal influence on density [19]; therefore, any differences in these damage calculations due to chromium content are expected to be minimal.

Ion beam and environmental conditions for the six implantations used are shown in table 2. The vacuum during implantation was maintained below $5 \times 10^{-4}$ Pa and there were no visible differences in the sample surfaces after implantation. $Fe^+$ ions were implanted successively with energies of 2MeV, 1 MeV and 0.5MeV to produce a roughly uniform damage profile with depth as shown in figure 1a. The beam current was maintained for each of the energies used during implantation by adjusting conditions at the source (for a coarse control) and adjusting the size of the beam scan area (for control to within 1μA/m$^2$). This enabled beam currents to be held constant for all energies, giving constant peak dose rates shown in figure 1b. The ion beam was focused to a diameter of 5mm and spread over the implantation area by a beam scanning; this produced a localised pulsing of the beam with increase in damage rate. The frame rate for scanning the beam over the samples was 4 kHz, resulting in a displacement damage with each frame of $1.5 \times 10^{-7}$ dpa for the high dose rate implantations and $7.5 \times 10^{-9}$ dpa for the low dose rate implantations. This is a small fraction of the dose required for cascade overlap, ~0.01dpa [14]; thus localised flux is not considered to influence dose rate. Essentially the radiation environment was held constant apart from varying dose rates of $3 \times 10^{-5}$ dpa/s and $6 \times 10^{-4}$ dpa/s and temperatures of 300, 400 and 500°C.

### 2.3	Nanoindentation

After ion implantation at 300°C, the region of the sample exposed to the beam was evident by a darker contrast in secondary electron imaging produced by both electrons and ions in the focused ion beam microscope (FIB). This enabled the FIB-milling of marker lines at the implanted / un-implanted boundary, visible under an optical microscope and which allowed nanoindentation testing of the implanted and un-implanted regions of the same grain in a given specimen. This contrast difference was not clearly visible in the materials subjected to irradiation at higher temperatures, and so the implanted / unimplanted boundary could not be so accurately delineated. In these specimens, nanoindentations in implanted and un-implanted regions were performed within different grains well away from the edge of the irradiated regions (two indentation arrays were made in different grains in both the irradiated and un-irradiated regions of each sample).

Nanoindentation was performed using a Nano Indenter-XP (MTS Systems Corp.) and NanoSuite 6.0 software. A diamond Berkovich tip was used with the Continuous Stiffness Measurement (CSM) method [20]; this allowed the continuous measurement of both hardness and elastic modulus as a function of indenter depth. The tip geometry was calibrated by indenting a fused silica reference sample with known mechanical properties. For each material condition, load-displacement data from 8 or more indent arrays were produced with a strain rate of 0.05 /s and analysed according to the method developed by Oliver and Pharr [20]. For samples irradiated at 300°C the crystallography at the position of each indent was analysed by Electron Back Scatter Diffraction (EBSD). Indents which fell outside the grain in which the main array was located or were within 20um of a grain boundary were omitted from the analysis.

## 2.4 Transmission Electron Microscopy (TEM)

Before irradiation, the as-received Fe 5%Cr alloy was mechanically polished to a thickness of approximately 150μm by SiC abrasive paper (final stage: FEPA P4000 SiC). 3mm diameter TEM discs were punched from this material and the discs were further mechanically polished to a thickness less than 100μm. All discs were heat treated at 800°C for 4 hours followed by slow cooling ( > 10 hours) in a vacuum <$10^{-4}$ Pa. After the treatment all specimens were examined by TEM and found to have a simple ferritic microstructure with grain size larger than 5 μm and low dislocation density. The TEM samples were irradiated alongside the bulk samples for mechanical testing, with the same conditions described in Section 2.2.

After irradiation, the foils were electro-polished for a few seconds in order to access the peak damage region. This was achieved by a bath polishing technique with accurate temporal control. The samples were then electro-polished into TEM thin foils using a Tenupol 5 Jet-electropolisher.
TEM observations were made using a Phillips CM20 microscope with an operating voltage of 200kV. TEM characterization techniques such as kinematical bright-field (KBF) and dynamical two-beam diffraction conditions were used to characterise radiation damage.

## 2.5    Atom probe Tomography

APT specimens were prepared from the ion-implanted bulk samples by focussed ion beam milling (FIB) using a Zeiss NVision 40 with a Kleindiek micromanipulator according to the 'lift-out' method outlined by Saxey et al. [21]. APT data acquisition was carried out using a Cameca LEAP 3000HR instrument operating in laser-pulsing mode. The specimen base temperature was maintained at -223°C (50K) with a laser energy of 0.5nJ. A 8µm laser spot size was used at a repetition rate of 200 kHz.

## 3    RESULTS

Figure 2 shows representative hardness versus indenter displacement data for implanted and unplanted regions in two different grains in Fe5%Cr samples irradiated at a temperature of 300°C. The increase in hardness due to radiation damage near the surface of the irradiated region is evident at small indenter displacements; at greater depths, the hardness of the irradiated region tends towards that of the un-irradiated bulk of the grain. The variation of hardness with composition, dose rate and temperature was studied using the mean hardness between indenter displacements of 50 and 200nm. The 200nm limit was chosen with reference to cross-sectional TEM observations of indent plastic zone sizes at various indentation depths, in a comparable irradiated material [22], so that the indentation plastic zone was wholly contained within the damage zone in tests on irradiated material. The 50nm limit was used to avoid errors due to tip bluntness and any possible surface contamination layers. Two grains in both irradiated and un-irradiated regions of each sample were tested; data for all samples are shown in figure 3.

The range of variation due to differences in crystallographic orientation of the grains in which the indentation arrays were made can be seen by comparing hardness data in un-irradiated areas (covering several grains) over all samples of the same composition. Average values for hardness (and standard deviation) for all tests in the un-irradiated material were 1.61 GPa (0.09 GPa) for pure Fe, 1.67 GPa (0.08 GPa) for Fe 5%Cr, 1.93 GPa (0.08 GPa) for Fe 10%Cr and 2.27 GPa (0.16 GPa) for Fe 14%Cr. Hardness data in the irradiated material also exhibited little scatter. Differences in hardness due to crystallography were found to be minimal compared to those caused by irradiation.

After irradiation at 300°C (figure 3a), irradiation hardening was clearly observed in all samples. The radiation hardening effect was greatest in Fe5%Cr subjected to the low dose rate. The difference in hardness between low and high dose rates decreases with increasing Cr content and is not evident in Fe14%Cr. There were no effects of dose rate on irradiation hardening in pure Fe. After irradiation at 400°C at both dose rates (figure 3b), irradiation hardening was clearly observed in all Fe-Cr alloys, but not in pure Fe. Similar to the irradiation at 300°C, irradiation hardening in Fe5%Cr was greater at the lower dose rate, and this dose rate effect decreased with increasing Cr content. Figure 4 shows the difference between the average hardness for all indentation tests between the un-irradiated and irradiated material for all irradiation temperatures and dose rates.

Figure 5 shows TEM micrographs of the radiation damage for Fe 5%Cr specimens irradiated at 300°C with the high dose rate (HDR) and low dose rate (LDR). The samples had a similar thickness of approximately 145nm measured by using a convergent beam technique and analyses were based on measurement of 542 loops (HDR) and 327 loops (LDR). Damage observed in all TEM specimens was in the form of dislocation loops with average diameters of 6.7 nm (HDR) and 6.5 nm (LDR) and corresponding densities of 16.1 x $10^{21}$/$m^3$ (HDR) and 9.72 x $10^{21}$/$m^3$ (LDR).

Specimens of Fe 5%Cr irradiated with the high and low dose rates at 400°C were then analyzed by APT. Table 3 shows the overall composition measurements for both types of alloy, including C and N interstitial impurities which may have been introduced during irradiation. Figure 6 shows the 3-D atom maps generated from the APT analysis with a corresponding proximity histogram showing the concentration of Cr from the core of the enriched region, outwards into the matrix. Segregation of Cr is clearly observed in the alloy irradiated with a low dose rate (Figure 6a). There are also traces of nitrogen in the regions of enrichment, but other elements do not show evidence of segregation.

The high concentration of Cr in the matrix made defining Cr segregation difficult, due to artefacts in the areas near major crystallographic poles in the APT data. The presence of N highlighted several areas of Cr segregation, thus the CrN ion counts were used to identify the regions with Cr enrichment.

Regions of Cr enrichment in the alloy irradiated at the low dose rate were roughly 5-10 nm in diameter and were measured to have Cr contents up to 15 at.%. The number density of Cr-enriched regions identified by APT is ~2 x $10^{-21}$ m$^{-3}$, which is just below the dislocation loop density of 9.72 x $10^{21}$ m$^{-3}$ measured by TEM. This suggests that these Cr clusters may be associated with the dislocation loops; however spatial correlation of the two defect types was not possible, and so this conclusion remains tentative. Some segregation of Cr was also observed in the alloy irradiated with a high dose rate, though the size and number density of enriched Cr regions was significantly smaller than the alloy irradiated with the lower dose rate.

## 4 DISCUSSION

The elementary defects which exist after a cascade event and their corresponding mobility are the primary factors which determine the evolution of radiation damage in a crystalline material. Molecular Dynamic (MD) simulations provide means to study picosecond cascade events and have indicated the surviving defect formation for various energies and temperatures in iron. Simulations of cascade damage in iron have been performed studying the effects of parameters such as the primary knock-on atom (PKA) energy and temperature of the target material [23-25]. Vacancies in iron have high migration energy relative to that of SIAs, thus many vacancies remain as point defects directly after a cascade event. Interstitials have higher mobility, resulting in the surviving fraction of interstitials forming clusters. Cascade damage in high energy cascades at a temperature of 600K consists of single vacancies or di-vacancies at the cascade core, with a 'shell' of several self interstitial atoms (SIA) and SIA clusters consisting of up to tens of SIAs [26].

TEM observations of iron and its alloys subjected to self ion-implantation have identified that the agglomeration of defects into visible clusters does not occur until doses at which cascade volumes begin to overlap, corresponding to ~0.01dpa [14]. At the relatively high dose rates corresponding to ion irradiation experiments, damage generated from a single cascade event remains in the form of point defects and clusters smaller than ~1nm (invisible in the TEM) until interaction with additional cascades. Assuming a cascade overlap dose of 0.01dpa and using damage calculations produced by TRIM, the average duration between cascade overlap in the experiments reported here was ~17 seconds for the high dose rate and ~5.5 minutes for the low dose rate. Therefore the observed differences in hardening between both dose rates indicate that defect migration and subsequent microstructural evolution occur over timescales greater than at least 17 seconds.

## 4.1 Defect Mobility

In these experiments, all materials were annealed producing large grains with a low initial dislocation density. The low density of defect sinks and the relatively high dose rates used suggest that all materials were likely to have been irradiated under a recombination-dominated regime. An increase in defect mobility with irradiation temperature is evident in all irradiated materials as a reduction in irradiation hardening caused by a higher fraction of defect recombination (as shown in figure 4). The effect of an increase in irradiation hardening in Fe-Cr alloys with decreasing dose rate is reduced with increasing Cr content. It is likely that the pinning of defects by Cr [13, 14, 27, 28] reduces defect mobility and suppresses the differences in the evolution of damage between the two dose rates. Thus, a 'dose rate effect' was not observed in Fe14%Cr irradiated at 300°C, but was observed after irradiation at the higher temperature of 400°C, which can be attributed to an increase in defect mobility with temperature. In Fe14%Cr irradiated at the low dose rate, the increased defect mobility at an irradiation temperature of 400°C resulted in slightly greater irradiation hardening than at 300°C. This was not observed in Fe14%Cr irradiated at 400°C with the high dose rate.

At an irradiation temperature of 300°C there was no effect of dose rate on hardening in pure Fe. Pure Fe irradiated at 400°C exhibited no irradiation hardening, which suggests that the majority of radiation induced defects may annihilate by recombination at 400°C in pure Fe. At an irradiation temperature of 500°C, negligible hardening was observed in all materials, which suggests that at this temperature the majority of defects annihilate by recombination in the Fe-Cr alloys also. Thus defects may overcome pinning by Cr at this temperature.

**4.2    Irradiation Hardening**

Irradiation hardening is seen for indentation depths less than 200nm, when the "plastic zone" of the indent is predominantly within the damaged layer [22]. Once the plastic zone expands beyond the damaged layer, the observed hardness tends towards that of the un-irradiated material.

A previous study has identified that plastic deformation in $Fe^+$ irradiated Fe-Cr alloys progresses via the annihilation of radiation induced defects by glissile dislocations, causing dislocation channelling [22]. This was associated with the annihilation of defects by reaction with glissile dislocations [29] and may result in a reduction in the observed irradiation hardening. At irradiation temperatures of 300°C and 400°C, irradiation at a lower dose rate resulted in an increase in irradiation hardening of many Fe-Cr alloys. Despite this, the radiation damage microstructure observable in the TEM had a lower dislocation loop density in Fe 5%Cr irradiated with the low dose rate, compared to the same alloy irradiated with the high dose rate. Thus, the observed increase in irradiation hardening at the lower dose rate is likely to be due to the evolution of defects more complex than simple dislocation loops, which more strongly resist penetration and annihilation by glissile dislocations. The Cr-rich regions observed to be present at a high density in the Fe 5%Cr alloy irradiated with the low dose rate are likely to present such barriers to glissile dislocations. This is similar to the significant hardening associated with Cr-rich precipitate formation in "475°C embrittlement" in steels with >12%Cr [30-32]. The contribution of Cr-rich precipitates to hardening is also consistent with the lack of a dose rate dependence for hardening in pure Fe, where no Cr is present.

In a recombination - dominant system, a lower dose rate results in a smaller density of defects per unit volume per unit time and interaction between radiation induced defects will occur less frequently. This is likely to result in a smaller density of extended stable defect nucleation points (e.g. clusters and dislocation loops), with a larger number of defects migrating to each point. This is likely to have caused the observed reduction in loop density with decreasing dose rate (also observed elsewhere [9]) and Cr precipitation caused by heightened rates of radiation induced precipitation (RIP). RIP may be suppressed by an increasing dose rate, due to heightened rates of recombination resulting in a higher density of mobile defects which migrate over smaller distances. This compares well to the suppression of swelling and creep with increasing dose rates observed in austenitic stainless steels [7, 8].

The reduction of an irradiation hardening dependence of dose rate with increasing Cr content may be the result of reduced mobility of radiation induced defects by pinning at Cr atoms in the lattice (see section 4.1). Cr pinning may suppress RIP in alloys with increasing Cr content. This dependence on defect mobility is consistent with the onset of a hardening dependence on dose rate in Fe14%Cr as irradiation temperatures rises from 300°C to 400°C.

### 4.3  Fe-Cr Phase Stability

The enthalpy of mixing in Fe-Cr alloys changes sign from a negative to a positive at a composition of 8-10%, although the exact composition where phase decomposition occurs (especially at low temperatures) is a subject of debate [33]. During irradiation at 300-400°C there should be no thermodynamic driving force for Cr segregation in the Fe5%Cr alloy and very little if any in the Fe10%Cr alloy. Hence, 475°C embrittlement only occurs in alloys with >12%Cr. Despite this, there have been a small number of experimental observations of Cr clustering and enrichment at dislocation loops in Fe-Cr alloys with ≤10%Cr [34-37]. The mechanisms which cause Cr precipitation in irradiated materials are not fully understood, however recent modelling has shown that Cr solubility in Fe changes under hydrostatic pressure in the lattice [38]. This may provide the driving force for precipitation/segregation at the stressed regions surrounding radiation induced defects such as dislocation loops within the lattice. Evidence of precipitation of Cr at dislocation loops has been identified in an Fe 9%Cr alloy irradiated at 673K to 1dpa, by the observation of moiré fringes within the loops in the TEM [39]. At present, it is not clear whether the Cr precipitates are associated with the dislocation loops and further work is required to understand this behaviour.

In the alloys with ≥10%Cr, alpha prime (α') precipitation may cause increased hardening "475°C embrittlement hardening" [31]. The Fe14%Cr sample irradiated at 400°C at the low dose rate exhibited a slight increase in hardness in the un-irradiated region. This sample was held at 400°C for approximately 11 hours during the implantation, which may have initiated the early formation of α', however further analysis is required to confirm this behaviour. The combination of RIP of Cr in alloys with a low Cr content, and α' formation in alloys with a high Cr content, may give rise to the observed minimum increase in ductile to brittle transition temperature in alloys containing approximately 9%Cr [40].

## 5    SUMMARY and CONCLUSIONS

For the Fe-Cr alloys investigated in this study, ion implantation with a relatively low dose rate caused significantly higher irradiation hardening compared with the same alloys irradiated with a higher dose rate. This effect appears to be associated with radiation-driven clustering of chromium atoms into Cr-rich precipitates, which may be associated with the dislocation loops resulting from displacement damage.

This variation of irradiation hardening in Fe-Cr alloys with dose rate emphasises that effects of irradiation on materials' microstructure and properties for a given total dose will be influenced by the characteristic dose rate of the radiation source used, and illustrates the possible complexities of the dependence of irradiation hardening on irradiation temperature, dose rate and alloy composition. The ion implantation dose rates used here ($3 \times 10^{-5}$ – $6 \times 10^{-4}$ dpa s$^{-1}$) are several orders of magnitude greater than dose rates typical of components in power plants ($10^{-12}$ – $10^{-7}$ dpa s$^{-1}$). Thus changes in microstructure and mechanical properties after ion implantation to a given dose and their dependence on alloy composition will not be simply related to the effects observed after irradiation to the same dose in nuclear reactors with a much slower dose rate. This highlights both the need for caution in application of ion-irradiation data to reactor materials, and the need for a better understanding of the basic processes involved, so as to improve the interpretation and correlation of such data.

**ACKNOWLEDGEMENTS**

The work reported here was partially supported by the Engineering and Physical Science Council (EPSRC) via a Programme Grant "Materials for Fusion and Fission Power", EP/H018921. C.D. Hardie thanks EPSRC and the Culham Centre for Fusion Energy (CCFE) for funding in the form of an Industrial CASE studentship. The authors thank the European Fusion Development Agreement (EFDA) for the provision of materials and the Surrey Ion Beam Centre, UK, for assistance with ion implantation, in particular the work of A. Cansell and Dr N.Peng. The authors gratefully acknowledge discussions with Prof. S Dudarev and colleagues at CCFE.

**TABLE CAPTIONS**

**Table 1 - Chemical analysis of the alloys after hot forming and in the as-delivered final metallurgical condition (in brackets)** [17]**.**

**Table 2 - Irradiation conditions used at Surrey Ion Beam Centre, UK.**

**Table 3 – Composition measured by APT of Fe 5%Cr samples after irradiation at 400°C with the low dose rate and high dose rate.**

**FIGURE CAPTIONS**

**Figure 1** - TRIM calculations showing displacement damage with implantation depth: (a) displacement damage level for all implantations and dose rates (b) damage rates for high dose rate (6 x $10^{-4}$ dpa/s) and low dose rate (5 x $10^{-5}$ dpa/s).

**Figure 2** - Average hardness versus indenter displacement into the test surface for indentation arrays in Fe5%Cr. Data are for the irradiated and un-irradiated regions of two grains in each of low and high dose rate samples. Error bars represent +/- one standard deviation.

**Figure 3** - Average hardness values for data produced between indentation depths of 50 and 200nm for irradiation temperatures of (a) 300°C, (b) 400°C and (c) 500°C. Data are for indentation arrays within the same grain at low

dose rate (LDR: solid diamond), high dose rate (HDR: solid square) and the un-irradiated counterpart (open symbols). Data for each %Cr are laterally spread for ease of visibility.

**Figure 4** - Average irradiation hardening (ΔGPa) for materials subjected to high dose rate (a) and low dose rate (b) for irradiation temperatures of 300°C (solid black), 400°C (solid grey) and 500°C (open symbols). The reference value (ΔH=0) is that for un-irradiated material of the same composition.

**Figure 5** –TEM kinematical bright-field (KBF) micrograph for the Fe5%Cr alloy irradiated at 300°C with low dose rate and high dose rate. Both micrographs were taken near the [001] zone axis with **g**=1-10

**Figure 6** – APT data for the Fe5%Cr alloy irradiated at 400°C with the low dose rate (a) and high dose rate (b). The figure includes an atom map showing Fe atoms and a 0.5at.% CrN isoconcentration surface (i) and proximity histograms showing the variation in composition from the centre of enrichment outwards into the matrix (ii). Centre of enrichment is defined as the region of highest CrN concentration.

**ADDITIONAL COLOUR FIGURE CAPTION**

**Figure 3** - Average hardness values for data produced between indentation depths of 50 and 200nm for irradiation temperatures of (a) 300°C, (b) 400°C and (c) 500°C. Data are for indentation arrays within the same grain at low

dose rate (LDR: solid blue), high dose rate (HDR: solid red) and the un-irradiated counterpart (open symbols). Data for each %Cr are laterally spread for ease of visibility.

# TABLES AND FIGURES

**Table 1 - Chemical analysis of the alloys after hot forming and in the as-delivered final metallurgical condition (in brackets)** [17]**.**

| Alloy | C wt ppm | S wt ppm | O wt ppm | N wt ppm | P wt ppm | Cr |
|---|---|---|---|---|---|---|
| Fe | 3 (4) | 2 (2) | 5 (4) | 2 (1) | <5 | <2 ppm |
| Fe 5%Cr | 3 (4) | 3 (3) | 4 (6) | 3 (2) | <5 | 5.40 wt% |
| Fe 10%Cr | 4 (4) | 6 (4) | 3 (4) | 3 (3) | <5 | 10.10 wt% |
| Fe 14%Cr | 4 (5) | 6 (7) | 4 (4) | 5 (5) | <10 | 14.25 wt% |

**Table 2.2 - Irradiation conditions used at Surrey Ion Beam Centre, UK.**

| Irradiation Temperature (°C) | Dose Rate (dpa/s) | Energy (MeV) | Dose (ions/m$^2$) | Dose Rate (ions/m$^2$/s) | Beam Current (μA/m$^2$) |
|---|---|---|---|---|---|
| 300, 400 and 500 | 6 x 10$^{-4}$ | 0.5 | 5 x 10$^{17}$ | 1.87 x 10$^{15}$ | 299 |
| | | 1 | 7.5 x 10$^{17}$ | 2.16 x 10$^{15}$ | 345 |
| | | 2 | 2.5 x 10$^{18}$ | 2.28 x 10$^{15}$ | 366 |
| | 3 x 10$^{-5}$ | 0.5 | 5 x 10$^{17}$ | 9.35 x 10$^{13}$ | 15 |
| | | 1 | 7.5 x 10$^{17}$ | 1.08 x 10$^{13}$ | 17.3 |
| | | 2 | 2.5 x 10$^{18}$ | 1.14 x 10$^{14}$ | 18.3 |

**Table 3 – Composition measured by APT of Fe 5%Cr samples after irradiation at 400°C with the low dose rate and high dose rate.**

| Element | Low dose rate (at.%) | High dose rate (at.%) |
|---|---|---|
| Fe | 94.74 | 94.7 |
| Cr | 5.26 | 5.31 |
| N | 0.045 | 0.02 |
| C | 0.02 | 0.02 |

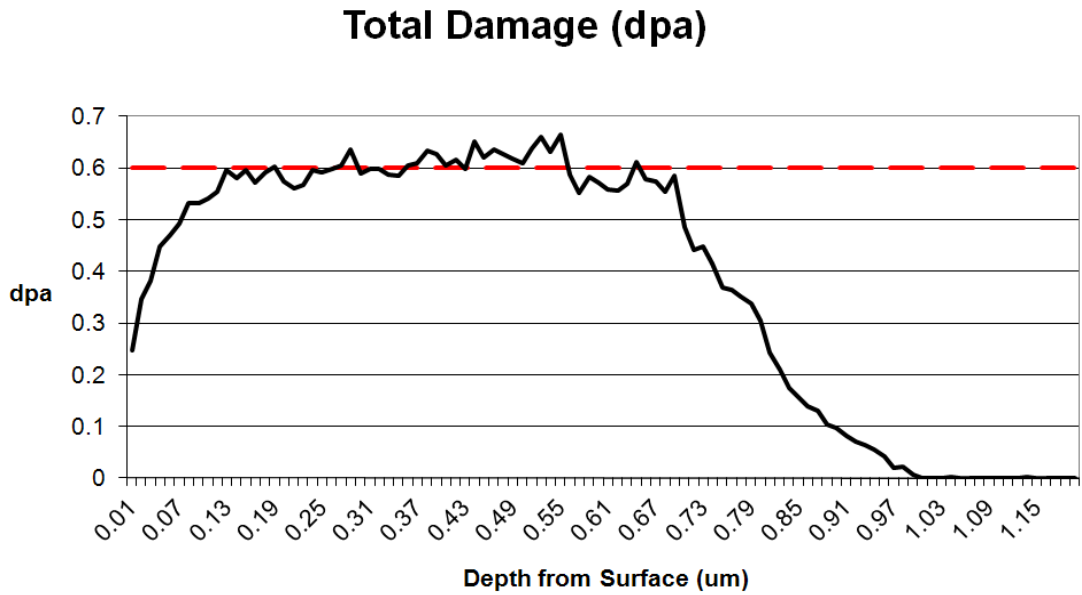

a.

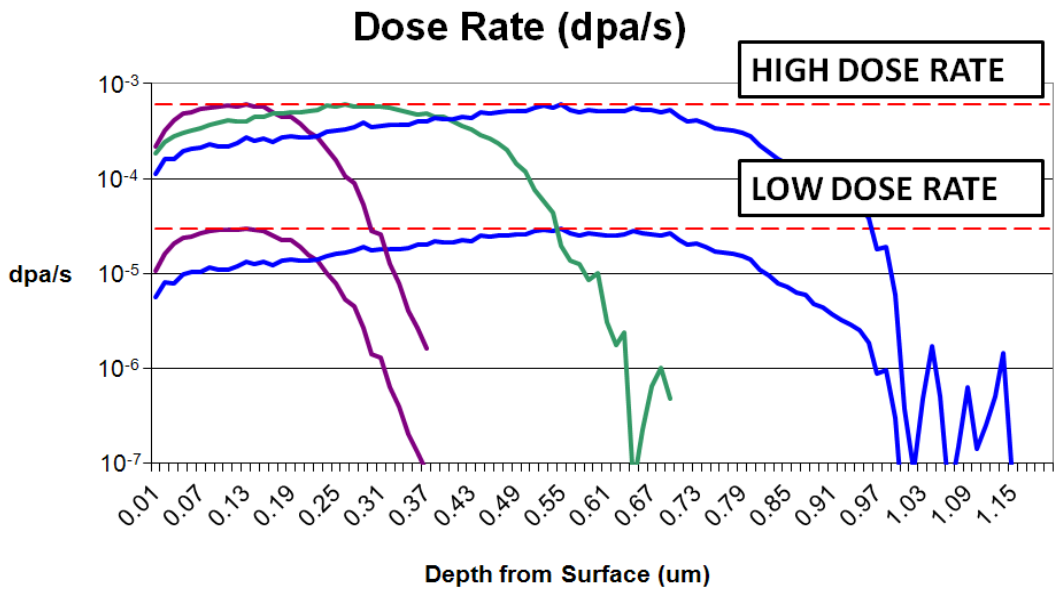

b.

**Figure 1** - TRIM calculations showing displacement damage with implantation depth: (a) displacement damage level for all implantations and dose rates (b) damage rates for high dose rate ($6 \times 10^{-4}$ dpa/s) and low dose rate ($5 \times 10^{-5}$ dpa/s).

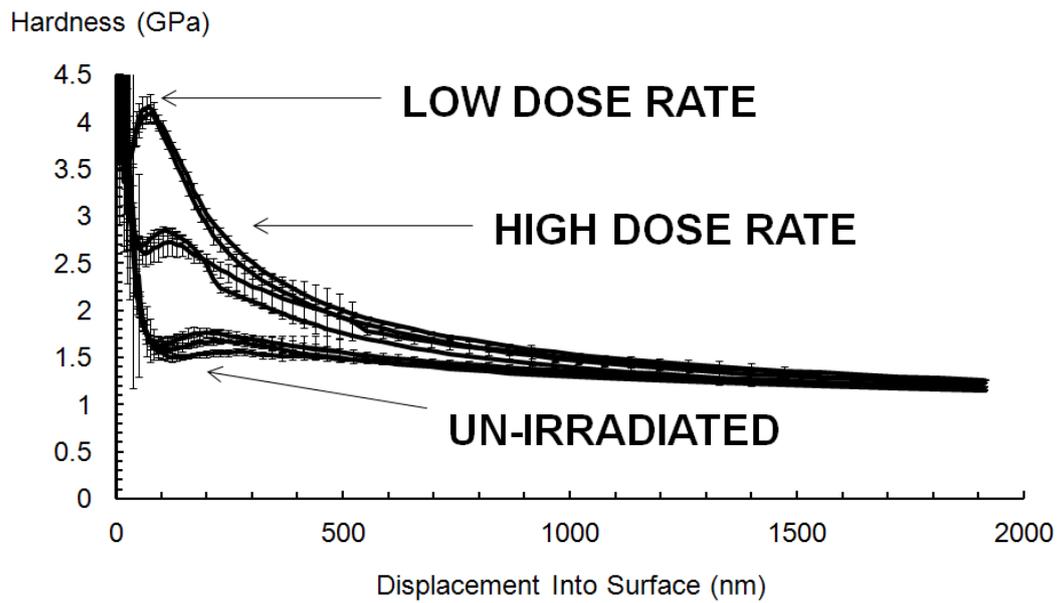

Figure 2 - Average hardness versus indenter displacement into the test surface for indentation arrays in Fe5%Cr. Data are for the irradiated and un-irradiated regions of two grains in each of low and high dose rate samples. Error bars represent +/- one standard deviation.

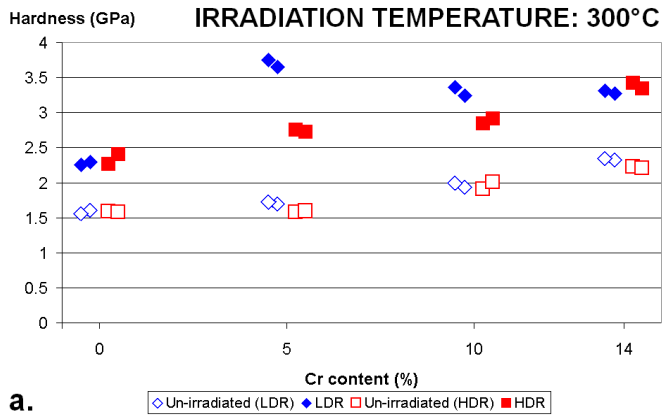

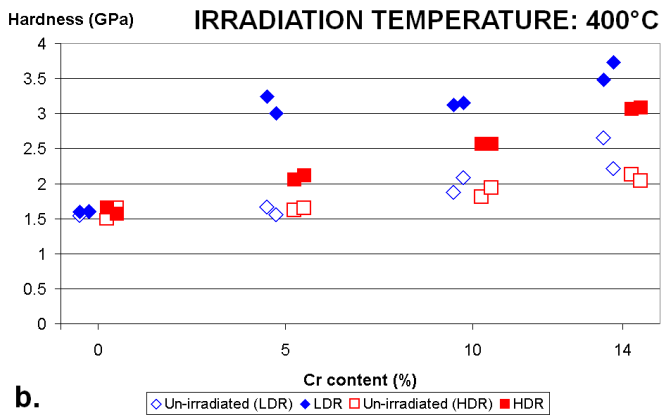

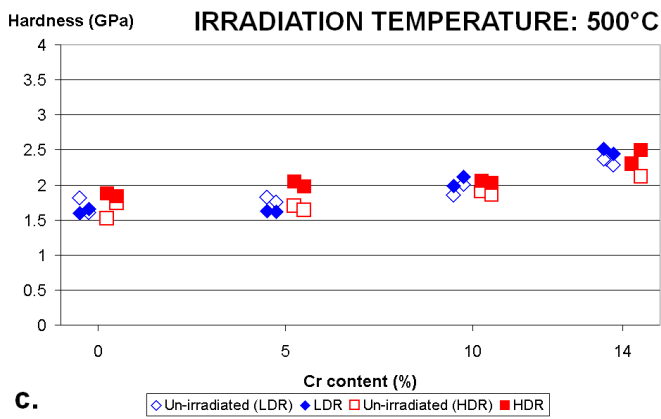

**Figure 3** - Average hardness values for data produced between indentation depths of 50 and 200nm for irradiation temperatures of (a) 300°C, (b) 400°C and (c) 500°C. Data are for indentation arrays within the same grain at low dose rate (LDR: solid blue), high dose rate (HDR: solid red) and the un-irradiated counterpart (open symbols). Data for each %Cr are laterally spread for ease of visibility.

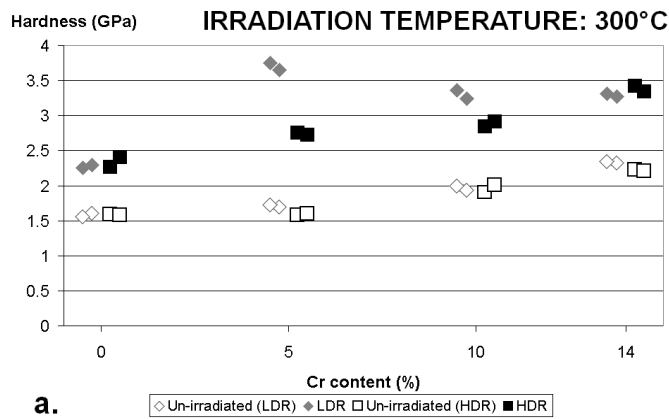
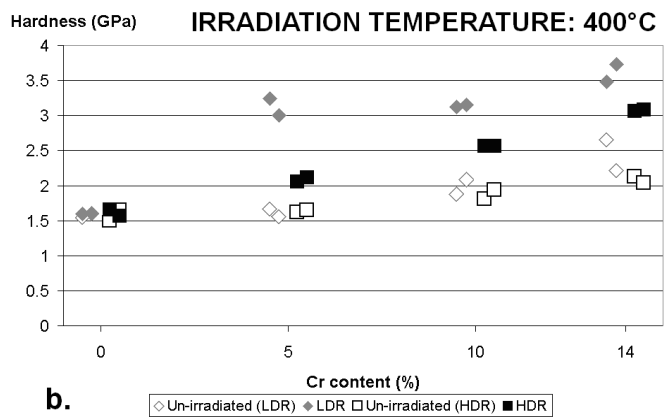
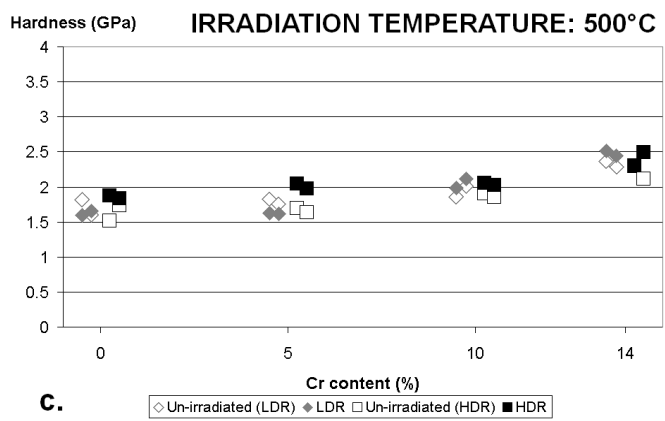

**Figure 3** - Average hardness values for data produced between indentation depths of 50 and 200nm for irradiation temperatures of (a) 300°C, (b) 400°C and (c) 500°C. Data are for indentation arrays within the same grain at low dose rate (LDR: solid diamond), high dose rate (HDR: solid square) and the un-irradiated counterpart (open symbols). Data for each %Cr are laterally spread for ease of visibility.

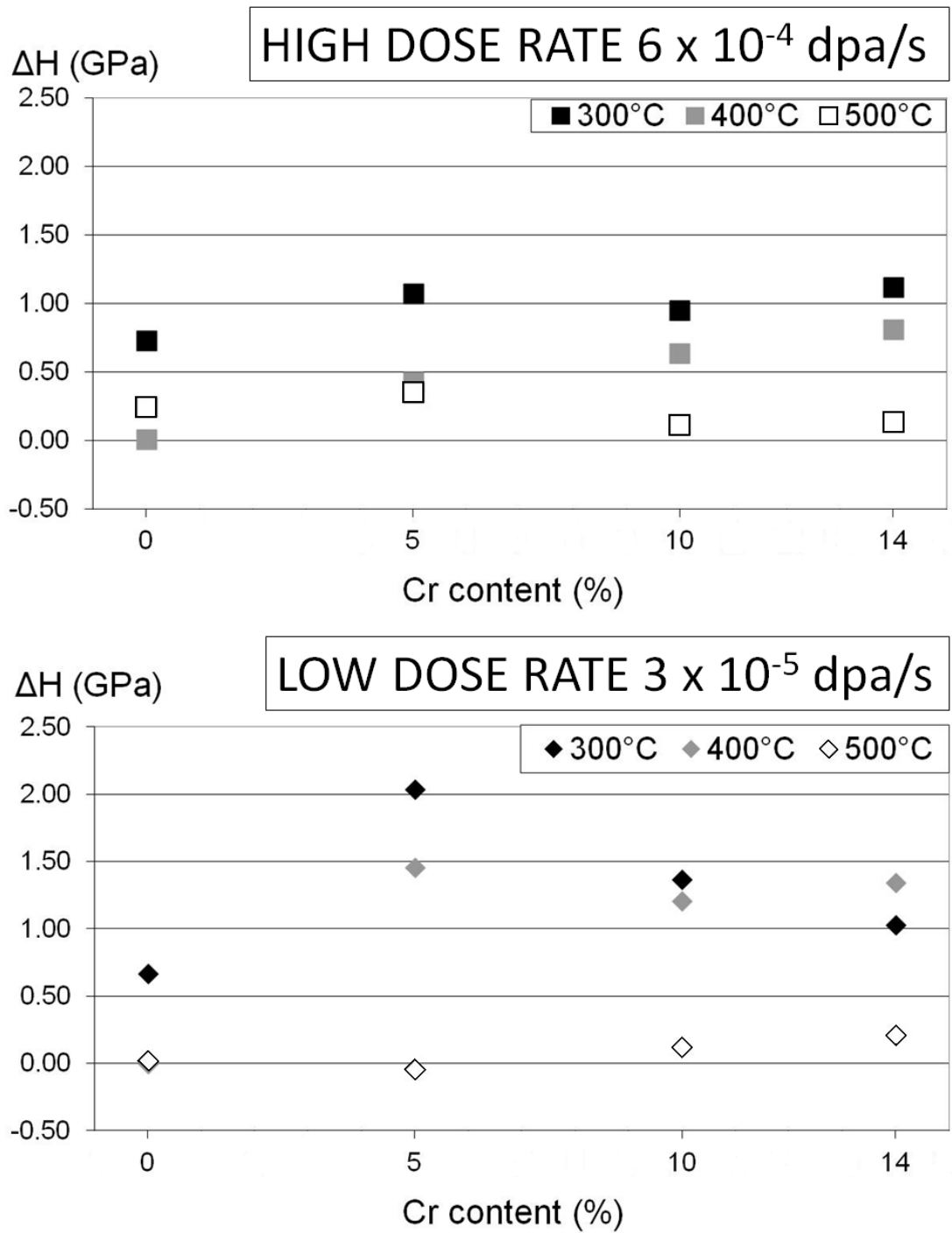

Figure 4 - Average irradiation hardening (ΔGPa) for materials subjected to high dose rate (a) and low dose rate (b) for irradiation temperatures of 300°C (solid black), 400°C (solid grey) and 500°C (open symbols). The reference value (ΔH=0) is that for un-irradiated material of the same composition.

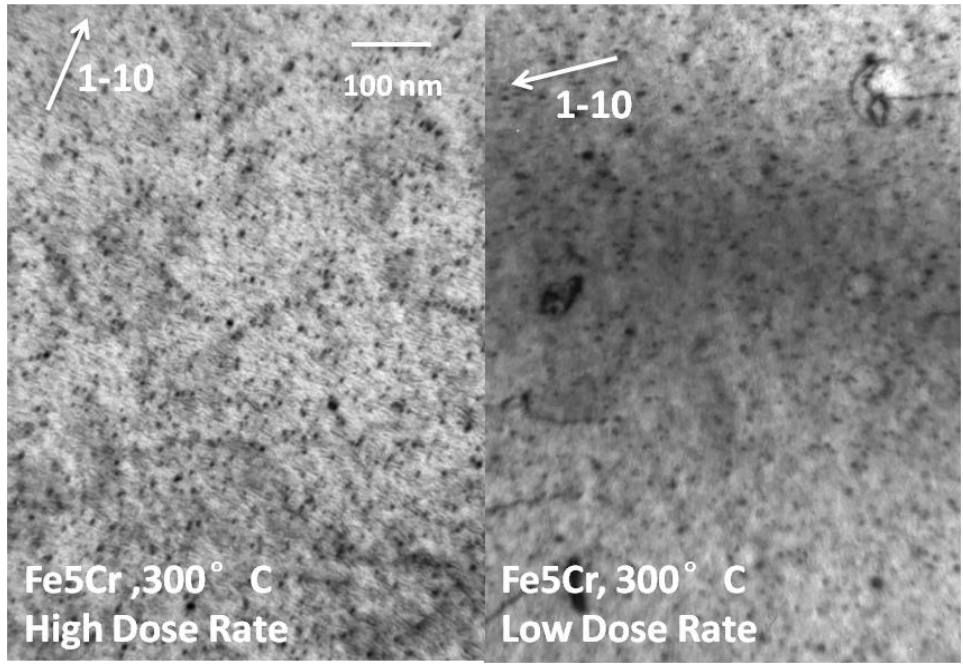

**Figure 5** –TEM kinematical bright-field (KBF) micrograph for the Fe5%Cr alloy irradiated at 300°C with low dose rate and high dose rate. Both micrographs were taken near the [001] zone axis with **g**=1-10

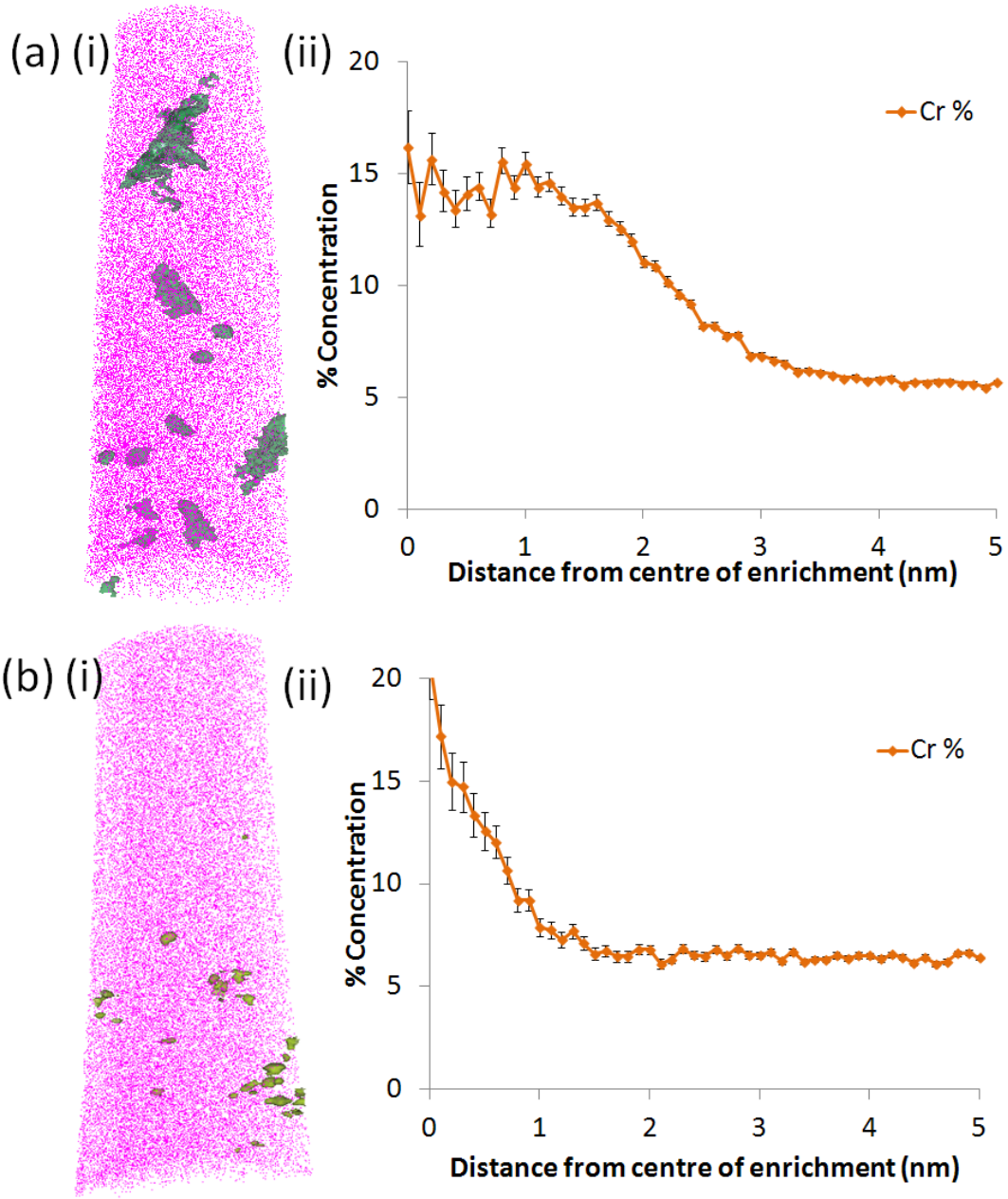

**Figure 6** – APT data for the Fe5%Cr alloy irradiated at 400°C with the low dose rate (a) and high dose rate (b). The figure includes an atom map showing Fe atoms and a 0.5at.% CrN isoconcentration surface (i) and proximity histograms showing the variation in composition from the centre of enrichment outwards into the matrix (ii). Centre of enrichment is defined as the region of highest CrN concentration.

Vitae

Christopher D Hardie

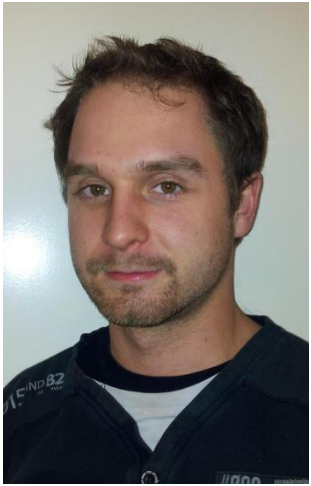

Christopher Hardie is a DPhil student in the Department of Materials at the University of Oxford. His research focuses on materials subjected to severe environments characteristic to existing and new nuclear power technologies. To date, his work has been centred on the use of heavy ion irradiation for the investigation of different radiation conditions and their effects on the mechanical properties of ferritic alloys.

Professor Steve G Roberts

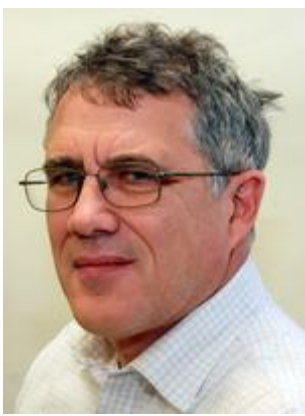

Professor Steve Roberts is a Professor of Materials at Oxford University, with over 30 years experience researching mechanical properties of materials, especially the links between dislocation behaviour, flow and fracture in ceramics, metals and alloys. He has recently developed new techniques to study mechanical properties at the micron scale, making direct measurements of yield, flow and fracture of thin films, and fracture strengths and stress corrosion cracking of individual grain boundaries. He leads the EPSRC funded programme "Materials for Fusion and Fission Power" and is co-director of the "Advanced Research" theme of the Bristol-Oxford Nuclear Research Centre.

Shuo Xu

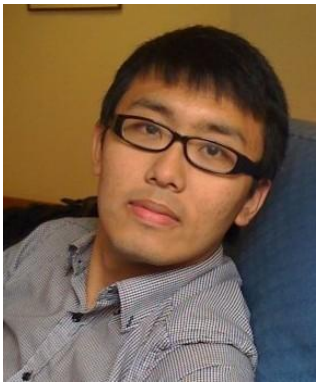

Shuo Xu is a DPhil student in the Department of Materials at the University of Oxford. His research focused on irradiation damage characterization for pure Fe and FeCr alloys using transmission electron microscopy. Specifically at the types of defects, the densities produced, and the ways that they interact with each other to change the mechanical properties of the materials.